\documentclass{PoS}

\usepackage{amsmath}
\usepackage{graphicx}
\usepackage{tikz}
\usepackage{slashed}
\usepackage{epstopdf}
\usepackage{mathrsfs}  
\usepackage{cite}

\title{Properties of Yang-Mills scattering forms}

\ShortTitle{Properties of Yang-Mills scattering forms}

\author{\speaker{Leonardo de la Cruz}\\
         Higgs Centre for Theoretical Physics, School of Physics and Astronomy,\\
         The University of Edinburgh\\
        E-mail: \email{leonardo.delacruz@ed.ac.uk}}

\author{Alexander Kniss\\
        PRISMA Cluster of Excellence, Johannes Gutenberg-Universit\"at Mainz\\
        E-mail: \email{akniss@students.uni-mainz.de}}
        
\author{Stefan Weinzierl\\
        PRISMA Cluster of Excellence, Johannes Gutenberg-Universit\"at Mainz\\
        E-mail: \email{weinzierl@uni-mainz.de}}

\abstract{In this talk we introduce the properties of scattering forms on  the compactified 
moduli space of Riemann spheres with $n$ marked points. These differential forms are $\text{PSL}(2,\mathbb{C})$ invariant, their intersection numbers correspond to scattering amplitudes as recently proposed by Mizera. All 
singularities are at the boundary of the moduli space  and  each singularity is logarithmic. In addition, each 
residue factorizes into two differential forms of lower points.}

\FullConference{Loops and Legs in Quantum Field Theory (LL2018)\\
		29 April 2018 - 04 May 2018\\
		St. Goar, Germany}

\begin{document}
\section{Introduction}
A fundamental property of scattering amplitudes is factorization. This property may be encoded in an auxiliary 
space such as the moduli space of Riemann spheres with $n$ marked points $\mathcal{M}_{0,n}$, which  is typical 
in string amplitudes and more recently appears in the Cachazo-He-Yuan (CHY) formalism \cite{Cachazo:2013gna, Cachazo:2013hca, Cachazo:2013iea}. In the CHY formalism amplitudes are evaluated on the solutions of the scattering equations which are points on $\mathcal{M}_{0,n}$. The CHY formula is a contour integral which schematically reads
\begin{align}
 \mathcal{A}_n( p, \varepsilon) = \mathrm{i} \oint_{\mathcal{O}} \; 
 I(z, p, \varepsilon) \; \mathrm{d} \Omega_{\text{CHY}},\qquad
f_i(z, p)\equiv\sum\limits_{\substack{j=1 \\ j\ne i}}^n \frac{ 2 p_i \cdot p_j}{z_i-z_j}=0,
\end{align}
\noindent where the contour encloses inequivalent solutions of the scattering equations $f_i(z, p)=0$. The integrand $I(z, p, \varepsilon)$ is  theory-dependent. 

A recent approach by Arkani-Hamed-Bai-He-Lam-Yan \cite{Arkani-Hamed:2017tmz, Arkani-Hamed:2017mur} suggests to rethink amplitudes directly the kinematic space. This is done by formulating amplitudes as differential forms in positive kinematic space. Differential forms may also be formulated in the auxiliary space and then mapped to amplitudes by reinterpreting them as intersection numbers\cite{Mizera:2017rqa}, where the ingredients
of these differential forms are the (half) integrands appearing in the CHY formula.    
In this talk we will introduce well defined tree-level scattering forms on the compactification of $\mathcal{M}_{0,n}$ for bi-adjoint scalar amplitudes and  Yang-Mills amplitudes. These forms satisfy properties that mimic the properties of the auxiliary space \cite{delaCruz:2017zqr}. 

\subsection{Results}
We define the cyclic and polarization scattering forms in terms of cyclic and polarization factors, respectively. In order to define them we first introduce a differential form
\begin{align}
 \frac{\mathrm{d}^n z}{ \mathrm{d} \omega},
\end{align}
\noindent where
\begin{align}
 \mathrm{d} \omega=(-1)^{p+q+r} 
\dfrac{\mathrm{d} z_p \mathrm{d}  z_q \mathrm{d} z_r}{(z_{p}-z_{q})(z_{q}-z_{r})(z_{r}-z_{p})}.
\end{align}
\noindent The cyclic (or Parke-Taylor) factor is defined by  
\begin{align}
 C(\sigma, z) =  \frac{1}{z_{\sigma_1 \sigma_2} z_{\sigma_2 \sigma_3}\cdots z_{\sigma_n \sigma_1}},
\end{align}
\noindent where $z_{ij}=z_i-z_j$. Finally, the polarization factor is constructed using numerators $N^{\text{BCJ}}_{\text{comb}}$  associated with comb diagrams (Fig.\ref{comb-diagrams})  of the BCJ decomposition of Yang-Mills amplitudes\footnote{See Section 2.2}. The polarization factor is defined by 
\begin{align}
 E(p, \varepsilon, z)= \sum\limits_{\kappa \in S_{n-2}^{(i, j)}} C(\kappa, z) N^{\text{BCJ}}_{\text{comb}} (\kappa),
\end{align}
\noindent  where  $i,j\in \{1,\dots,n\}$ and $\kappa$ is a permutation of $\{1, \dots, n\}$ with $\kappa_1=i$ and $\kappa_n=j$. 
 These factors allow us to define scattering forms on the full space $\bar{\mathcal{M}}_{0, n}(\mathbb{C})$(the compactification) as follows 
\begin{align}
 \Omega^{\text{cyclic}}_{\text{scattering}}(\sigma, z) \equiv  C(\sigma, z) \frac{\mathrm{d}^n z}{ \mathrm{d} \omega}, \nonumber \qquad 
\Omega^{\text{pol}}_{\text{scattering}}(p, \varepsilon, z)  \equiv  E(p,\varepsilon, z) \frac{\mathrm{d}^n z}{ \mathrm{d} \omega}. 
\end{align}
\noindent These forms satisfy the properties:
\begin{enumerate}
 \item $\text{PSL}(2, \mathbb{C} )$ invariance.
 \item Twisted intersection numbers give amplitudes \cite{Mizera:2017rqa}, e.g., for Yang-Mills primitive
 amplitudes
 \begin{align}
 A_n(\sigma, p, \varepsilon) = ( \Omega^{\text{cyclic}}_{\text{scattering}}(\sigma, z), \Omega^{\text{pol}}_{\text{scattering}}(p, \varepsilon, z))_{\eta} \sim \text{CHY}. \nonumber
 \end{align}
 \item Singularities are on $\bar{\mathcal{M}}_{0,n} \setminus \mathcal{M}_{0,n}$.
 \item Logarithmic singularities.
 \item Residues factorize  into two scattering forms of lower points.
\end{enumerate}
Some of these properties are similar to those  of  scattering forms on  \emph{positive} kinematic space 
\cite{Arkani-Hamed:2017tmz, Arkani-Hamed:2017mur}. E.g., residues of scattering forms on kinematic space factorize into forms of lower points as well.

\section{Known results}
\subsection{Moduli space of genus zero curves}
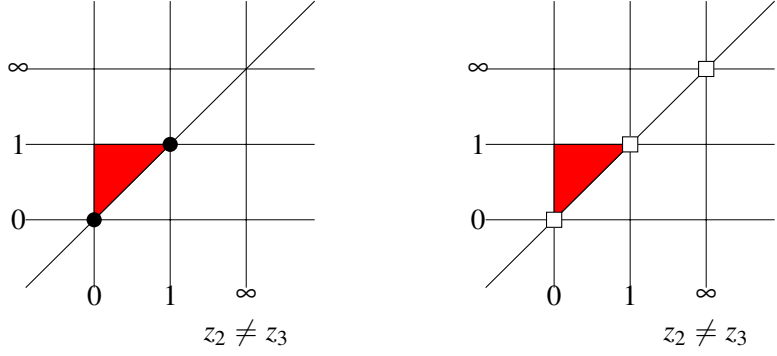
\begin{figure}
\centering
\begin{minipage}{0.4\textwidth} 
\begin{tikzpicture}
\draw[step=1cm,black,very thin] (-1.9,-1.9) grid (1.9,1.9);
\draw (-1.9,-1.9) -- (1.9, 1.9);
\filldraw[fill=red] (0,0)  -- (-1,0)  -- (-1,-1)   -- cycle;
\node[text width=0.5cm, text centered ] at (-2,-1) {$0$};
\node[text width=0.5cm, text centered ] at (-2,0) {$1$};
\node[text width=0.5cm, text centered ] at (-2,1) {$\infty$};
\node[text width=0.5cm, text centered ] at (-1,-2) {$0$};
\node[text width=0.5cm, text centered ] at (-0,-2) {$1$};
\node[text width=0.5cm, text centered ] at (+1,-2) {$\infty$};
\node[text width=2.0cm, text centered ] at (+1,-2.5) {$z_2\ne z_3$};
\fill[black] (-1,-1) circle (0.1cm);
\fill[black] (0,0) circle (0.1cm);
\end{tikzpicture}
\end{minipage}%
 \begin{minipage}{0.4\textwidth}
\begin{tikzpicture}
\draw[step=1cm,black,very thin] (-1.9,-1.9) grid (1.9,1.9);
\draw (-1.9,-1.9) -- (1.9, 1.9);
\filldraw[fill=red] (0,0)  -- (-1,0)  -- (-1,-1)   -- cycle;
\node[text width=0.5cm, text centered] at (-2,-1) {$0$};
\node[text width=0.5cm, text centered] at (-2,0) {$1$};
\node[text width=0.5cm, text centered] at (-2,1) {$\infty$};
\node[text width=0.5cm, text centered] at (-1,-2) {$0$};
\node[text width=0.5cm, text centered] at (-0,-2) {$1$};
\node[text width=0.5cm, text centered] at (+1,-2) {$\infty$};
\node[text width=2.0cm, text centered] at (+1,-2.5) {$z_2\ne z_3$};
\filldraw[fill=white] (-1.1,-1.1) rectangle (-0.9,-0.9);
\filldraw[fill=white] (-0.1,-0.1) rectangle (0.1,0.1);
\filldraw[fill=white] (0.9, 0.9) rectangle (1.1,1.1); 
\end{tikzpicture}
\end{minipage}
\caption{$\mathcal{M}_{0,5}(\mathbb{R})$ is the complement of five lines(left). Bounded region on $\mathcal{M}_{0,5}(\mathbb{R})$ (left, red). The space  $\bar{\mathcal{M}}_{0,5}(\mathbb{R})$, obtained from $\mathcal{M}_{0,5}(\mathbb{R})$ by blowing up the points $(0,0)$, $(1,1)$, and 
$(\infty, \infty)$(right). After blowing up these points, the colored region becomes a pentagon.}
\label{moduli-space-5}
\end{figure}

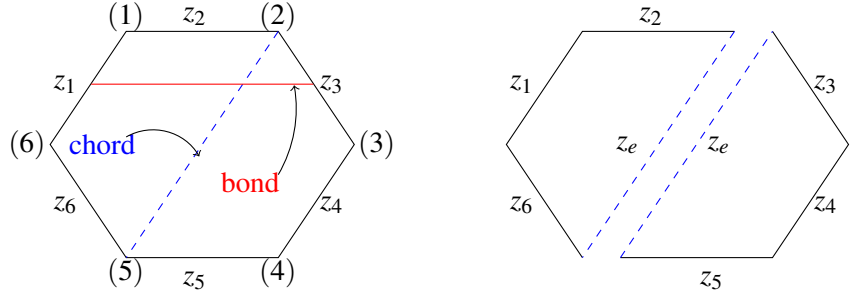
\begin{figure}
\centering
 \begin{tikzpicture}
\draw (0.5, 0)--(1.5, 1.5)--(3.5,1.5)--(4.5,0)--(3.5,-1.5)--(1.5,-1.5)--cycle;
\node[text width=0.5cm] at (0.2,0) {$(6)$};\node[text width=0.5cm] at (1.5,1.7) {$(1)$};\node[text width=0.5cm] at (3.5,1.7) {$(2)$};
\node[text width=0.5cm] at (4.8,0) {$(3)$}; \node[text width=0.5cm] at (3.5,-1.7) {$(4)$};  \node[text width=0.5cm] at (1.5,-1.7) {$(5)$};
\node[text width=0.5cm] at (0.8,0.8) {$z_1$}; \node[text width=0.5cm] at (2.5,1.7) {$z_2$};\node[text width=0.5cm] at (4.3, 0.8) {$z_3$};
\node[text width=0.5cm] at (4.3,-0.8) {$z_4$}; \node[text width=0.5cm] at (2.5,-1.8) {$z_5$};  \node[text width=0.5cm] at (0.8,-0.8) {$z_6$};
\draw[dashed, blue] (3.5,1.5)--(1.5, -1.5); 
\node[text width=1.5cm, blue] at (1.5, 0) {chord};
\draw [->] (1.5,0.1) arc (120:30:20pt);
\node[text width=1.5cm, red] at (3.5, -0.5) {bond};
\draw[red] (1.04, 0.8)--(3.96, 0.8);
\draw [->] (3.5, -0.4) arc (-30:10:50pt); 
\draw (7.5,-1.5)--(6.5, 0)--(7.5, 1.5)--(9.5,1.5);
\draw[dashed, blue] (9.5,1.5)--(7.5, -1.5); 
\draw (10.0, 1.5)--(11.0,0)--(10.0,-1.5)--(8.0,-1.5);
\draw[dashed, blue] (8.0,-1.5)--(10.0, 1.5); 
\node[text width=0.5cm] at (6.8,0.8) {$z_1$}; \node[text width=0.5cm] at (8.5,1.7) {$z_2$}; \node[text width=0.5cm] at (6.8, -0.8) {$z_6$};
\node[text width=0.5cm] at (8.2, 0.) {$z_e$};
\node[text width=0.5cm] at (9.4, 0.) {$z_e$};\node[text width=0.5cm] at (10.8,0.8) {$z_3$}; \node[text width=0.5cm] at (10.8, -0.8) {$z_4$};
\node[text width=0.5cm] at (9.3,-1.8) {$z_5$};  
\end{tikzpicture}
\caption{Dihedral structure and factorization of the $n$-gon along $u_{2, 5}$. On the $n$-gon sides are associated with the coordinates,  chords join two vertices, and a bond connects two sides of the $n$-gon (left).   
The dihedral extension satisfy the property that a divisor is the product of spaces of the same type (right)}
\label{chords-bonds-factor}
\end{figure}

The moduli space of genus zero curves is an $n-3$ dimensional variety defined by   
\begin{align}
 \mathcal{M}_{0,n} (\mathbb{C})=\{(z_2, \dots, z_{n-2})\in \mathbb{C}^{n-3}: z_i\ne z_j, z_i \ne 0, z_i\ne 1 \}.
\end{align}
\noindent In order to visualize the space let us consider the real part for  $n=5$. In this case, we have the space depicted in Fig.\ref{moduli-space-5}. Consider the region colored in red bounded by $z_2=0$, $z_3=1$, and $z_2=z_3$. These lines do not cross normally at the points $(0, 0)$ and $(1,1)$ (three divisors meet at these points). In order to fix the  situation we can blow up these points. Following \cite{Brown:2009qja}, we consider dihedral structures $(\pi, z)$ to achieve this. A dihedral structure can be represented  by the identification of the coordinates $z$ with the sides of an $n$-gon labeled by the permutation $\pi$ (See Fig.\ref{chords-bonds-factor}).     
Given a dihedral structure we define chords as lines joining two vertices of the labeled $n$-gon and we assign
the cross ratios 

\begin{align}
u_{i,j}= \frac{(z_i-z_{j+1})(z_{i+1}-z_j)}{(z_i-z_j)(z_{i+1}-z_{j+1})}.
\end{align}

\noindent These cross ratios define coordinates of a new space called the dihedral extension   $\mathcal{M}^{\pi}_{0,n}$ of $\mathcal{M}_{0,n}$. By gluing the $(n-1)!/2$ dihedral structures we obtain the Deligne-Mumford-Knudsen compatification\footnote{We have $\mathcal{M}_{0,n}=\mathcal{M}_{0,z}$.} \cite{Knudsen:1969,Knudsen:1976, Knudsen:1983a, Knudsen:1983b}
\begin{align}
 \bar{\mathcal{M}}_{0,z}= \bigcup\limits_{\pi} \mathcal{M}^{\pi}_{0,z}.
\end{align}
\noindent The boundaries  $\mathcal{M}^{\pi}_{0,z}\backslash \mathcal{M}_{0,z}$ now cross normally and thus by constructing the dihedral extension we have blown up the original space. 

\subsection{Color-kinematics duality}

\begin{figure}
\centering
\includegraphics[keepaspectratio = true,width=0.45\textwidth]{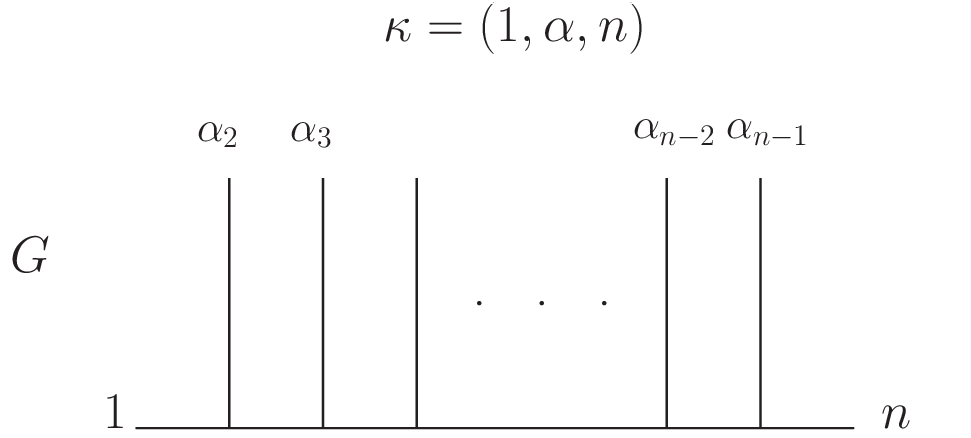} 
\caption{Comb diagrams $G$ with the standard ordering  $\kappa_1=1$ and $\kappa_n=n$.}
\label{comb-diagrams}
\end{figure}

A well known way of separating the group information from the kinematic information of a pure Yang-Mills amplitude is the color decomposition (see \cite{Weinzierl:2016bus} for a review) 
\begin{align}
 \mathcal{A}_n(p, \varepsilon)= g^{n-2} \sum\limits_{\sigma \in S_n/Z_n} 2\, \text{Tr} (T^{a_{\sigma(1)}} \cdots
 T^{a_{\sigma(n)}}) A_n(\sigma, p, \varepsilon).
\end{align}
\noindent Similarly, one can show that amplitudes can be decomposed  as
\begin{align}
 \mathcal{A}_n(p, \varepsilon)= \mathrm{i} \ g^{n-2} \sum\limits_{\text{trivalent graphs } G} \frac{C(G)N^{\text{BCJ}}(G)}{D(G)},
\end{align}
\noindent where the BCJ numerators $N^{\text{BCJ}}(G)$ satisfy antisymmetry and Jacobi-like identities whenever the color factors do. This concept is known as the color-kinematics duality \cite{Bern:2008qj, Bern:2010ue, Bern:2010yg}.  Using the properties of the numerators  one  can show that the BCJ numerators can be decomposed in terms only of multi-peripheral (comb) diagrams(Fig.\ref{comb-diagrams}). To these diagrams we associate the BCJ numerators 
\begin{align}
N^{\text{BCJ}}_{\text{comb}}(\kappa), 
\end{align}
\noindent where $\kappa$ labels the external ordering of the diagram. In order to compute these numerators we write and effective lagrangian \cite{Tolotti:2013caa}
\begin{align}
\mathscr{L}= \frac{1}{2 g^2} \sum\limits_{n=2}^\infty \mathscr{L}^{(n)}. 
\end{align}
\noindent For example 
  \begin{align}
  \mathscr{L}^{(4)}=& - g^{\mu_1\mu_3} g^{\mu_2\mu_4} g_{\nu_1 \nu_2} 
  \frac{\partial_{12}^{\nu_1} \partial_{34}^{\nu_2}}{
  \Box_{12}}\text{Tr} [\mathsf{A}_{\mu_1}, \mathsf{A}_{\mu_2}] [\mathsf{A}_{\mu_3}, \mathsf{A}_{\mu_4}].
 \end{align}
 \noindent The above term leads to Feynman rules which can be used to compute the numerators. This is equivalent 
 to the introduction of auxiliary tensor particles to eliminate 4-gluon vertex \cite{Draggiotis:1998gr, Duhr:2006iq, Weinzierl:2016bus}. 

\section{Scattering forms}
Let us now introduce the scattering forms. Without loss of generality let us take  $\pi =(1, \dots, n)$. Notice that the dihedral structure defines a patch of $\bar{\mathcal{M}}_{0,n}$ and that a given permutation $\sigma$ contains the information about the external ordering of the amplitude, so we have two relevant permutations for our problem. 
\subsection{Cyclic scattering forms}
\begin{figure}
\centering
\begin{tikzpicture}
\draw (0.5, 0)--(1.5, 1.5)--(3.5,1.5)--(4.5,0)--(3.5,-1.5)--(1.5,-1.5)--cycle;
\node[text width=0.5cm] at (0.2,0) {$(6)$};\node[text width=0.5cm] at (1.5,1.7) {$(1)$};\node[text width=0.5cm] at (3.5,1.7) {$(2)$};
\node[text width=0.5cm] at (4.8,0) {$(3)$}; \node[text width=0.5cm] at (3.5,-1.7) {$(4)$};  \node[text width=0.5cm] at (1.5,-1.7) {$(5)$};
\node[text width=0.5cm] at (0.8,0.8) {$z_1$}; \node[text width=0.5cm] at (2.5,1.7) {$z_2$};\node[text width=0.5cm] at (4.3, 0.8) {$z_3$};
\node[text width=0.5cm] at (4.3 ,-0.8) {$z_4$}; \node[text width=0.5cm] at (2.5,-1.7) {$z_5$};  \node[text width=0.5cm] at (0.8,-0.8) {$z_6$};
\draw[blue,dashed] (4.5,0.0)--(0.5, 0.0); 
\draw[red] (1.04, 0.8)--(3.96, 0.8)--(2.5,1.5)--(3.96,-0.8)--(2.5,-1.5)--(1.04,-0.8)--cycle;
\end{tikzpicture} 
\caption{Bond diagram for the permutation $\sigma=(1,3,2,4,5,6)$ and dihedral structure $\pi=(1,2,3,4,5,6)$}
\label{bonds}
\end{figure}
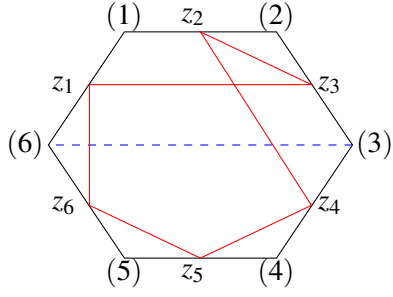

The scattering form is then defined as 
\begin{align}
 \Omega^{\text{cyclic}}_{\text{scattering}}(\sigma, z) \equiv  C(\sigma, z) \frac{\mathrm{d}^n z}{ \mathrm{d} \omega}.
\end{align}
\noindent For example, in the particular case where $\pi=\sigma$ the scattering form in dihedral coordinates may be written as 
\begin{align}
 \Omega^{\text{cyclic}}_{\text{scattering}}(\pi, u)= \prod\limits_{j=2}^{n-2} \frac{1}{u_{j,n}(u_{j,n}-1)} \mathrm{d}^{n-3} u.
\end{align}
 In order to establish the properties of the cyclic form we will to introduce a useful construction. A bond connects two edges of the $n$-gon associated with a dihedral  structure $\pi$ (Fig.2). For a given chord $(i_0, n)$ we  say that $\sigma$ and $\pi$ are equivalent if exactly two bonds cross the chord $(i_0, n)$. In Fig.\ref{bonds} we have e.g. $(1,3,2,4,5,6) \sim_{(3,6)} (1,2,3,4,5,6)$.
\noindent This construction allows us to answer the question of what happens with $\Omega^{\text{cyclic}}_{\text{scattering}}(\sigma, z)$ when $u_{i_0, n}$ goes to zero. The answer is that the number of bonds counts the powers of $u_{i_0, n}$ in $\Omega^{\text{cyclic}}_{\text{scattering}}(\sigma, z)$. Therefore the logarithmic singularities and residue factorization will  follow from this analysis. 

Let us now determine the properties of the cyclic scattering form.
\begin{itemize}
 \item  Under the transformation $z\rightarrow g(z)=(az+b)/(cz+d)$ the cyclic 
 factor and measure transform as
  \begin{align}
  C(\sigma, z)& \rightarrow \prod\limits_{j=1}^n (c z_j +d)^{2} C(\sigma, z),\\
  \frac{\mathrm{d}^n z}{\mathrm{d}\omega} &\rightarrow \prod\limits_{j=1}^n (c z_j +d)^{-2} \frac{\mathrm{d}^n z}{\mathrm{d}\omega},
 \end{align}
\noindent respectively. Hence $\text{PSL}(2, \mathbb{C})$ follows.
 \item Intersection numbers. Bi-adjoint scalars may be written as \emph{twisted} intersection numbers of two cyclic scattering forms, say, with  orderings $\sigma$ and $\tilde{\sigma}$. The twist is defined by 
  \begin{align}
  \eta= \sum\limits_{i=1}^n f_i (z, p) \mathrm{d}z_i.
 \end{align}
\noindent This twist makes the intersection number of two $n-3$ differential forms  equivalent to the CHY formula\footnote{The formal statements is that the amplitude is the twisted intersection number of two cocycles, twisted by $\eta$. \cite{Mizera:2017cqs}}. 
 \item The cyclic factor $C(\sigma, z)$ is singular when  $z_{\sigma_i}=z_{\sigma_{i+1}}$. These points are on the divisor
 $\bar{\mathcal{M}}_{0,n}\setminus \mathcal{M}_{0.n}$.
 \item Analyzing the equivalence between $\pi$ and $\sigma$, we obtain a factor of $u_{i_0,n}^{1-i_0}$ from the 
 cyclic factor and a factor of $u_{i_0,n}^{i_0-2}$ from the measure when  $\pi$ and $\sigma$ are equivalent, i.e., when exactly two  bonds cross the chord $(i_0, n)$. In contrast, we obtain fewer powers when $\pi$ and $\sigma$ are not equivalent. Hence singularities of the cyclic scattering form are logarithmic. 
 \begin{align}
\sigma \sim_{(i_0,n)} \pi & \qquad u_{i_0,n}^{1-i_0} \Big(u_{i_0,n}^{i_0-2} \Big),\nonumber\\
\sigma \slashed{\sim}_{(i_0,n)} \pi & \qquad \text{fewer powers}\nonumber
\end{align}

\item A similar analysis tell us that the residue at $u_{i_0, n}=0$ is zero if $\sigma$ and $\pi$ are not equivalent and it has a single pole otherwise. Denoting the hypersurface $u_{i_0, n}=0$ by $Y$, we have
\begin{equation}
\text{Res}_{Y} \Omega^{\text{cyclic}}_{\text{scattering}}(\sigma, z)= \begin{cases}
                                                              (-1)^{i_0-1} \Omega^{\text{cyclic}}_{\text{scattering}}(\sigma', z)
                                                              \wedge \Omega^{\text{cyclic}}_{\text{scattering}}(\sigma'', z), & \sigma \sim_{(i_0, n)} \pi,\\
                                                              0,&  \text{otherwise}
                                                             \end{cases}.
                                                             \end{equation} 
\end{itemize}
\noindent Cyclic scattering forms have also been studied in \cite{Mizera:2017cqs}  and 
\cite{arkani-hamed:2017a, bai:2017}.

\subsection{Polarization forms}
The polarization scattering form is defined by 
\begin{align}
 \Omega^{\text{pol}}_{\text{scattering}}(p,  \varepsilon, z)\equiv E(p,  \varepsilon, z)  \frac{\mathrm{d}^n z}{\mathrm{d}\omega} =\sum\limits_{\kappa\in S_{n-2}^{(i,j)}}
 C(\kappa, z) N^{\text{BCJ}}_{\text{comb}}(\kappa) \frac{\mathrm{d}^n z}{\mathrm{d}\omega},
\end{align}
\noindent where the sum runs over all permutations $\kappa$ with $\kappa_1=i$ and $\kappa_n=j$ fixed\footnote{This choice is arbitrary and it can be shown that the polarization factor is permutation invariant \cite{delaCruz:2017zqr}.}. This form contains three main ingredients:
\begin{enumerate}
 \item A cyclic factor $C(\kappa, z)$.
 \item A BCJ numerator associated with comb diagrams $N^{\text{BCJ}}_{\text{comb}}(\kappa)$.
 \item The invariant measure.
\end{enumerate}
\noindent In general the polarization factor 
\begin{align}
 E(p,\varepsilon)=\sum\limits_{\kappa\in S_{n-2}^{(i,j)}} C(\kappa, z) N^{\text{BCJ}}_{\text{comb}}(\kappa)
\end{align}
\noindent differs from the reduced Pfaffian of the CHY formalism.  The reduced Pfaffian has been extensively studied
in the literature \cite{Du:2013sha, Litsey:2013jfa, Lam:2016tlk, Bjerrum-Bohr:2016axv, Huang:2017ydz, Du:2017kpo,
Gao:2017dek, Chen:2016fgi, Chen:2017edo, Chen:2017bug}. The reduced Pfaffian 
 \begin{align}
\frac{(-1)^{i+j}}{2 z_{ij}} \text{Pf} \Psi^{ij}_{ij}, \qquad 1< i,j< n,
\end{align}

\noindent is independent of the choice of $i, j$ on the support of the scattering equations. In contrast, our formula is defined on the the full $\bar{\mathcal{M}}_{0,n}$ and coincides with the reduced Pfaffian in the subvariety defined by the scattering equations. Notice that the first and third factors
gives us some of the required properties of the polarization form due to the appearance of the cyclic form.   
However,  its definition on the full $\bar{\mathcal{M}}_{0,n}$ requires that, in general, polarizations are not transverse and momenta to be off-shell.

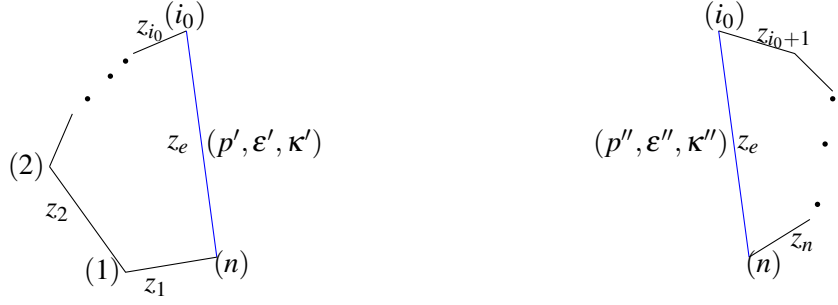
\begin{figure}
\centering
\begin{tikzpicture}
\draw (0.2, 0)--(0.5, 0.7);
\draw (1.3, 1.5)--(2.0, 1.8);
\draw[blue] (2.0, 1.8)--(2.4, -1.2);
\draw (2.4, -1.2)--(1.2, -1.4)--(0.2, 0);
\fill (0.7, 0.9) circle  [radius=1pt];
\fill (1.0, 1.2) circle  [radius=1pt];
\fill (1.2, 1.4) circle  [radius=1pt];
\node[text width=0.5cm] at (-0.1,0) {$(2)$};\node[text width=0.6cm, centered] at (2.0,2.0) {$(i_0)$};\node[text width=0.5cm] at (2.6, -1.3) {$(n)$};
\node[text width=0.5cm] at (0.9,-1.4) {$(1)$};
\node[text width=1.0cm, centered] at (9.3,2.0) {$(i_0)$};\node[text width=0.5cm] at (9.6, -1.3) {$(n)$};
\node[text width=0.5cm] at (2.0, 0.3) {$z_e$};\node[text width=0.5cm] at (9.5, 0.3) {$z_e$};\node[text width=0.5cm] at (1.6, 1.8) {$z_{i_0}$};  
\node[text width=0.5cm] at (1.7, -1.6) {$z_1$};\node[text width=0.5cm] at (0.4, -0.6) {$z_2$};\node[text width=0.5cm] at (10.2, -1.0) {$z_n$};
\node[text width=1.0cm, centered] at (10., 1.7) {$z_{i_0+1}$};
\node[text width=1.5cm] at (3.0, 0.3) {$(p', \varepsilon', \kappa')$};\node[text width=1.5cm] at (8.1, 0.3) {$(p'', \varepsilon'', \kappa'')$};
\draw[blue] (9.4, -1.2)--(9.0, 1.8);
\draw (9.0, 1.8)--(10.0, 1.5)--(10.5, 1.0);
\fill (10.5, 0.9) circle  [radius=1pt];
\fill (10.4, 0.3) circle  [radius=1pt];
\fill (10.3, -0.5) circle  [radius=1pt];
\draw (10.2, -0.7)--(9.4, -1.2);
\end{tikzpicture}
\caption{Factorization of data}
\label{factorization-data}
\end{figure}

In addition, we should define what factorization of numerators mean. Since the numerators depend on kinematic data and the orderings $\kappa$, we should find a definition which performs the factorization of data (Fig.\ref{factorization-data}).
\noindent Hence, for each $n$-gon we should have the data:

\begin{minipage}[t]{.45\textwidth}
\raggedright
\begin{itemize}
 \item $\varepsilon'=(\varepsilon_1, \varepsilon_2, \dots, \varepsilon_{i_0},\varepsilon_e)$,
 \item $p'=(p_1,  p_2, \dots, p_{i_0},p_q)$ ,
 \item $\kappa'=(1, {\kappa}_2',\dots, {\kappa}_{i_0}', e)$.
\end{itemize}
\end{minipage}
\begin{minipage}[t]{.45\textwidth}
\raggedleft
\begin{itemize}
 \item $\varepsilon''=(\varepsilon^{\ast}_e, \varepsilon_{i_0+1}, \dots, \varepsilon_{n})$,
  \item $p''=(\overline{p_q}, p_{i_0+1}, p_{i_0+2},  \dots, p_n)$,
   \item $\kappa''=(e, {\kappa}_{i_0+1}'',\dots, {\kappa}_{n-1}'', n)$.
\end{itemize}
\end{minipage}

\noindent Factorization of data introduces new off-shell momenta $p_q$ and $\overline{p}_q$. The sum over physical polarizations gives us a $4\times 4$ matrix (in Lorenz indices) of rank 2, which we supplement with two 
unphysical polarizations, such that 
\begin{align}
 \sum\limits_{\lambda} (\varepsilon_\mu^\lambda)^{\ast} \varepsilon_\nu^\lambda=-g_{\mu\nu}.
 \end{align}
  \noindent Similarly for the auxiliary particles, e.g., for the tensor particle 
  \begin{align}
\sum\limits_{\lambda} (\varepsilon_{\mu\nu}^\lambda)^{\ast} \varepsilon_{\rho\sigma}^\lambda=& -\frac{1}{2}
  p^2 (g_{\mu\rho} g_{\nu\rho}-g_{\mu\sigma}g_{\nu\rho}). 
 \end{align}
\noindent  With these definitions, the factorization of numerators reads
  \begin{align}
  N(G)= \sum\limits_{f, \lambda} N(G_1) N(G_2),
 \end{align}
 \noindent where the sum runs over particles and polarizations. Therefore the polarization factor  in terms
 of BCJ numerators  gives a good definition of a polarization factor. It is permutation invariant, 
 its dependence on $C(\kappa, z)$ implies properties 1,3,4  and it reproduces the CHY formula for 
 pure Yang-Mills amplitudes  $\mathrm{i} \oint \mathrm{d} \Omega_{\text{CHY}} C(\sigma, z) E(p, \varepsilon, z)$ 
 for on-shell momenta, physical polarizations and on the subvariety defined by the scattering equations. Hence the scattering form  satisfies property 2 as expected. 
 
Let us now sketch the proof of the factorization property. First, we have numerator factorization, i.e.,  
\begin{align}
  N^{\text{BCJ}}_{\text{comb}}((\kappa', \kappa''))= \sum\limits_{f, \lambda} 
  N^{\text{BCJ}}_{\text{comb}}(\kappa') N^{\text{BCJ}}_{\text{comb}}( \kappa''). \label{num-fac}
\end{align}
\noindent On the other hand, the residues of a cyclic scattering form factorize, i.e.,   
\begin{align}
 \text{Res}_{Y} \Omega^{\text{cyclic}}_{\text{scattering}}(\sigma, z)= (-1)^{i_0-1}  \Omega^{\text{cyclic}}_{\text{scattering}}(\kappa', z)  \wedge  \Omega^{\text{cyclic}}_{\text{scattering}}(\kappa'', z). \label{res-fac}
\end{align}   
 \noindent Therefore combining Eqs.\eqref{num-fac}-\eqref{res-fac}, we have
  \begin{align}
 \text{Res}_{Y} \Omega^{\text{pol}}_{\text{scattering}}(p, \varepsilon, z)=& \nonumber \\
 (-1)^{i_0-1}&  \sum\limits_{f, \lambda}  \sum\limits_{\kappa', \kappa''}  N^{\text{BCJ}}_{\text{comb}}(\kappa')
  \Omega^{\text{cyclic}}_{\text{scattering}}(\kappa', z)  \wedge N^{\text{BCJ}}_{\text{comb}}(\kappa'') \Omega^{\text{cyclic}}_{\text{scattering}}(\kappa'', z),
\end{align}
\noindent i.e., 
\begin{align}
 \text{Res}_{Y} \Omega^{\text{pol}}_{\text{scattering}}(p, \varepsilon, z)=\sum\limits_{f, \lambda} (-1)^{i_0-1}  \Omega^{\text{pol}}_{\text{scattering}}(p', \varepsilon', z) \wedge
  \Omega^{\text{pol}}_{\text{scattering}}(p'', \varepsilon'', z).
\end{align}

\section{Summary and Outlook}
In this talk we have presented the properties of scattering forms $\Omega^{\text{cyclic}}_{\text{scattering}}$ and $\Omega^{\text{pol}}_{\text{scattering}}$ defined on the  full $(n-3)$ dimensional space $\bar{\mathcal{M}}_{0,n}$ away from the solutions of scattering  equations. The factorization property of the polarization form forced us to introduce some non-physical polarizations. Properties 1-5 builds a bridge from differential forms from the  CHY formalism to ideas involving associahedra on kinematic and auxiliary space \cite{Arkani-Hamed:2017mur}. We have now a  clear geometric picture of   tree-level amplitudes  within bi-adjoint, Yang-Mills and gravity for any number of external particles. It would be interesting  to extend these ideas to theories which admit a CHY representation \cite{Cachazo:2014xea, delaCruz:2015raa,Cachazo:2014nsa}. It would be interesting to explore these ideas at loop level.

\end{document}